\documentclass[twocolumn,showpacs,a4,prapplied,superscriptaddress]{revtex4} 
\usepackage{graphicx}%
\usepackage{color}
\usepackage{dcolumn} 
\usepackage{amsmath} 
\usepackage{amssymb} 
\usepackage{bm} 
\usepackage{hyperref}
\usepackage{tabularx}
\hypersetup{colorlinks,%
						citecolor=blue,%
						filecolor=blue,%
						linkcolor=blue,%
						urlcolor=blue,%
						pdftex}
\fontfamily{ptm}
\begin{document}
\preprint{SIZE AND TEMPERATURE DEPENDENCE OF WATER DIFFUSION IN CNTS.TEX}  
\title{The Peculiar Size and Temperature Dependence of Water Diffusion in Carbon Nanotubes studied with 2D NMR Diffusion-Relaxation (\(D-T_{2eff}\)) Spectroscopy}

\author{L. Gkoura}
\affiliation{Institute of Nanoscience and Nanotechnology, NCSR Demokritos, 15310 Aghia Paraskevi, Attiki, Greece}
\author{G. Diamantopoulos}
\affiliation{Institute of Nanoscience and Nanotechnology, NCSR Demokritos, 15310 Aghia Paraskevi, Attiki, Greece}
\affiliation{Department of Physics, Khalifa University of Science and Technology, 127788, Abu Dhabi, UAE}
\author{M. Fardis}
\affiliation{Institute of Nanoscience and Nanotechnology, NCSR Demokritos, 15310 Aghia Paraskevi, Attiki, Greece}
\author{D. Homouz}
\affiliation{Department of Physics, Khalifa University of Science and Technology, 127788, Abu Dhabi, UAE}
\affiliation{Department of Physics, University of Houston, Houston TX, USA}
\affiliation{Center for Theoretical Biological Physics, Rice University, Houston TX, USA}
\author{S. Alhassan}
\affiliation{Department of Chemical Engineering, Khalifa University of Science and Technology, 127788, Abu Dhabi, UAE}
\author{M. Beazi-Katsioti}
\affiliation{School of Chemical Engineering, National Technical University of Athens, Athens, 15780 Zografou, Athens, Greece}
\author{M. Karagianni}
\affiliation{Institute of Nanoscience and Nanotechnology, NCSR Demokritos, 15310 Aghia Paraskevi, Attiki, Greece}
\author{A. Anastasiou}
\affiliation{Institute of Nanoscience and Nanotechnology, NCSR Demokritos, 15310 Aghia Paraskevi, Attiki, Greece}
\author{G. Romanos}
\affiliation{Institute of Nanoscience and Nanotechnology, NCSR Demokritos, 15310 Aghia Paraskevi, Attiki, Greece}
\author{J. Hassan}
\email[Corresponding author: ]{jamal.hassan@ku.ac.ae }
\affiliation{Department of Physics, Khalifa University of Science and Technology, 127788, Abu Dhabi, UAE}
\author{G. Papavassiliou}
\email[Corresponding author: ]{g.papavassiliou@inn.demokritos.gr}
\affiliation{Institute of Nanoscience and Nanotechnology, NCSR Demokritos, 15310 Aghia Paraskevi, Attiki, Greece}

\date{\today}

\begin{abstract}
It is well known that water inside hydrophobic nano-channels diffuses faster than bulk water. Recent theoretical studies have shown that this enhancement depends on the size of the hydrophobic nanochannels. However, experimental evidence of this dependence is lacking. Here, by combining two-dimensional Nuclear Magnetic Resonance (NMR) diffusion-relaxation (\(D-T_{2eff}\)) spectroscopy in the stray field of a superconducting magnet, and Molecular Dynamics (MD) simulations, we analyze the size dependence of water dynamics inside carbon nanotubes (CNTs) of different diameters ($1.1$ nm to $6.0$ nm), in the temperature range of $265$ K to $305$ K. Depending on the CNTs diameter, the nanotube water is shown to resolve in two or more tubular components acquiring different self-diffusion coefficients. Most notable, a favourable CNTs diameter range (\(3.0-4.5\) nm) is experimentally verified for the first time, in which water molecule dynamics at the centre of the CNTs exhibit distinctly non-Arrhenius behaviour, characterized by ultrafast diffusion and extraordinary fragility, a result of significant importance in the efforts to understand water behaviour in hydrophobic nanochannels.
\end{abstract}
\maketitle

\section{Introduction}
The study of water diffusion inside Carbon Nanotubes (CNTs) has attracted great interdisciplinary interest as conduit for understanding nanofluidic properties in a variety of nanoporous systems having potential in many applications, such as  water treatment technologies\cite{Das2014}, drug delivery\cite{Ketabi2017,Zhang2011}, intracellular solute transport control \cite{Costa2016}, and energy storage systems \cite{Ma2008,Lee2002}. Theoretical methods, mostly Molecular Dynamics (MD) simulations\cite{Ketabi2017,Alexiadis2008,Werder2003}, have been used to investigate the structure and dynamics of water molecules inside CNTs. A major outcome of these studies is that water molecules tend to stratify in coaxial tubular sheets inside CNT channels\cite{Alexiadis2008,Bordin2013,Alexiadis2008a}. In certain CNT sizes, nanotubular water diffuses faster than bulk water \cite{Alexiadis2008,Barati2011,Holt2006} upon confinement. This fast water motion has been explained by several authors as due to H-bond modifications in the hydrophobic nanochannels, or due to geometrical constraints and curvature induced incommensurability between the water and the CNT walls \cite{Alexiadis2008}. 
From the experimental point of view, a great number of methods, such as infrared spectroscopy \cite{Chen1998}, Raman spectroscopy \cite{Kukovecz2002,Campidelli2007}, thermogravimetric analysis \cite{Landi2005}, Transmission Electron Microscopy (TEM) \cite{Bonifazi2006,Gogotsi2000,Gogotsi2001,Naguib2004,Ye2004}, X-ray Compton scattering \cite{Reiter2013,Reiter2016}, and Nuclear Magnetic Resonance (NMR) \cite{Perez2005,Urban2005,Blackburn2006,Abou-Hamad2011} have been widely used in the study of molecular confinement and transport through the CNT channels \cite{Hassan2016}. However, until now there is scarce experimental evidence at molecular scale, regarding the way water organizes and diffuses inside the CNTs, and the way these properties vary as a function of the channel size and temperature. 
These challenges are associated with a fundamental problem in the physics of soft matter, which is still not well understood, i.e. the microscopic origin of the temperature dependence of the structural and dynamic properties, such as structural relaxation times and transport coefficients, of confined liquids (see for example ref. \cite{Bohmer1994}). The manifestation of a non-Arrhenius behavior in the temperature dependence of both the translational and rotational dynamics of liquid water has led Angel to introduce the concept of fragility, a useful classification of liquids along a strong and fragile scale \cite{Bohmer1994}. According to this conception, a pure Arrhenius behavior classifies a strong liquid whereas a non-Arrhenius one signifies a fragile behavior. This is demonstrated for example in the supercooled state of water where an increase of the apparent activation energy is observed upon cooling, noticeable even at room temperature. Water confined in nanotubes exhibits similar phenomena as described in this work. 
In the above context, NMR is an important noninvasive tool with atomic scale resolution for studying water-surface and nano-confined water systems. Standard NMR experiments typically include longitudinal \(T_1\), transverse \(T_2\) relaxation times, self-diffusion coefficient $D$, and line-shape measurements. In a recent survey \cite{Hassan2016} of NMR studies on the water dynamics inside CNTs, it is revealed that the majority of the published reports \cite{Matsuda2006,Kyakuno2011,Das2010,Ghosh2004,Sekhaneha2006} focused on $^1$H-water NMR-lineshape versus temperature in order to establish the freezing point of the confined water. However, any change in the water structure and dynamics induced by the nano-confinement is expected to be also reflected in the $^1$H NMR \(T_1\), \(T_2\) and $D$ values of the water molecules. Such measurements for water-in CNTs have been rarely published \cite{Liu2014,Wang2008}. It is noticed that until now, the temperature dependence of the self-diffusion coefficient $D$ of water in CNTs has been examined mainly by using neutron scattering techniques, focused on one or two CNT diameters \cite{Chu2007,Mamontov2006,Briganti2017}. Furthermore, in the temperature range investigated in the present work, the difference between the Arrhenius and non-Arrhenius behavior could not be appreciated by neutron scattering methods \cite{Briganti2017}.

In this paper we utilize two-dimensional $^1$H NMR diffusion-relaxation correlation spectroscopy (\(D-T_{2eff}\)) to study water confinement in CNTs as a function of size ($1.1$ nm- $6.0$ nm) and temperature ($265$ K-$305$ K) \cite{Hassan2018,Phillips2005,Berendsen1987,Kimmich1991}. In addition, MD simulations on the same systems were carried out in order to acquire the way that water is organized inside the CNTs, and compare theoretical with experimentall $D$ values. Notably, in the CNTs diameter range $3.0$ nm - $4.5$ nm, water is shown to split into coaxial water tubular sheets (WTS), with the central one obtaining an order of magnitude higher $D$, in comparison to the outer WTS close to the CNT wall. This result is not foreseen by any theoretical calculation method performed until now. Even more, water molecules along the CNTs axis show remarkable deviation from the Arrhenius temperature dependence; a very fragile, almost liquid-like axial water component, persisting even at very low temperatures is formed, with fragility sufficiently higher than that of the bulk water. 

\section{Results and discussion}

\begin{figure}[t]
\includegraphics[width=0.5\textwidth, keepaspectratio]{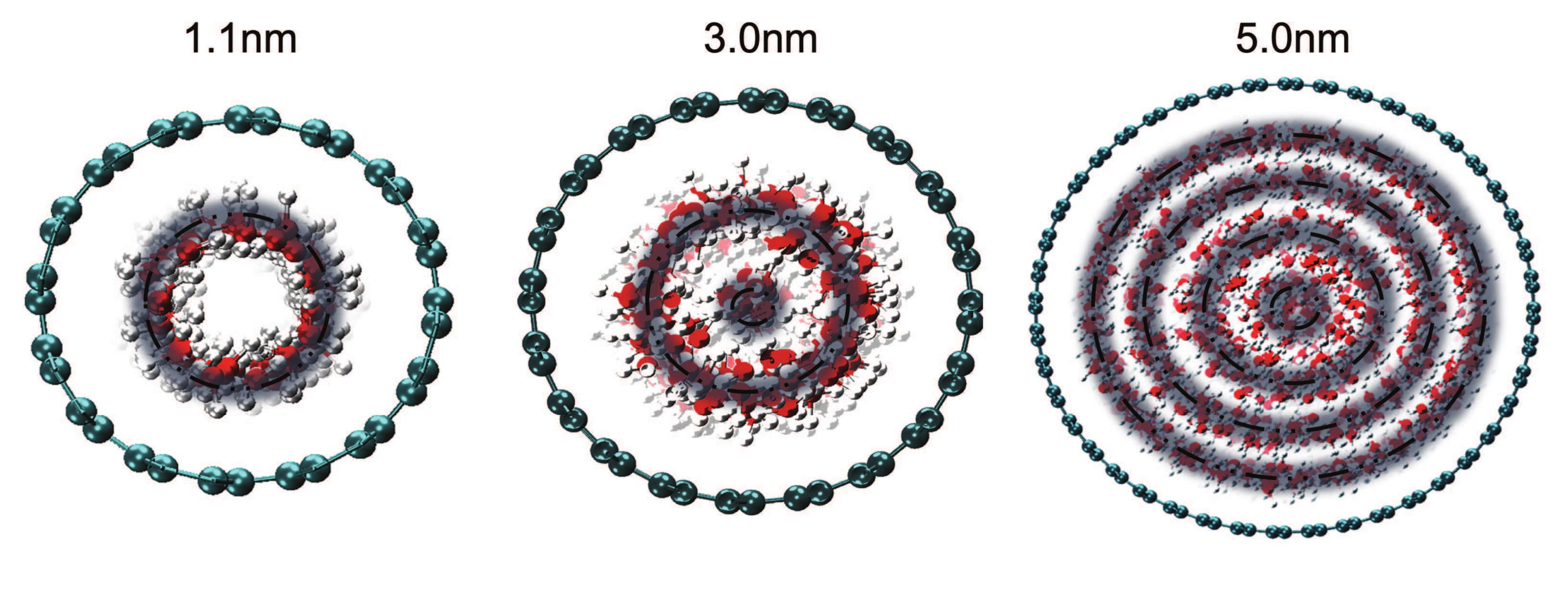}
\caption{\label{Fig1} Snapshots from MD simulations of water molecules arrangements inside CNT nanotubes of different sizes (green: carbon atoms of the CNT wall, red: Oxygen atoms white: Hydrogen atoms and black shaded-circles represent different water layers).}
\end{figure}

The water structure inside the CNTs varies upon increasing the CNTs diameter; it forms a $1D$ chain of water molecules in ultranarrow single walled CNTs (\(d<1\) nm), and it is organized into coaxial WTS as the CNTs diameter increases. Figure ~\ref{Fig1}, shows three MD simulation snapshots of water configuration at room temperature inside CNTs with diameters $1.1$ nm, $3.0$ nm, and $5.0$ nm. 

\indent

In all three cases, water molecules inside CNT channels are shown to arrange in concentric WTS (black circles in the snapshot), in agreement with previous publications (see for example Alexiadis et al. \cite{Alexiadis2008a} and references therein).  The number of WTS that can be accommodated inside CNT channels depends on the size of the CNTs and on the Oxygen-Oxygen as well as Oxygen-Carbon interactions \cite{Alexiadis2008a}. It is furthermore observed that the stratified water arrangement into WTS becomes denser by increasing the CNT size and gradually the dynamics of water molecules approach that of bulk water \cite{Alexiadis2008a}. 

In order to verify experimentally the role of the CNTs diameter on the water structure and dynamics, we conducted $2D$ $^1$H NMR \(D-T_{2eff}\) measurements of water in various CNT sizes, ranging from $1.1$ nm to $6.0$ nm, and in the temperature range $265$K to $305$K. Experiments were performed in the stray field of a superconductive magnet with a constant magnetic field gradient $g = 34.7$ T/m, at $^1$H NMR frequency of $101.324$ MHz. It is noticed that in a constant strong magnetic field gradient, as in our case, the CPMG spin-echo decay curves, assigned to the \(T_2\)-axis \cite{Hassan2018,Hurlimann2002}, decay with an effective \(T_{2eff}\), which is sufficiently shorter than  the intrinsic \(T_2\) in the absence of a magnetic field gradient. For example, \(T_{2eff}\) of bulk water was found to be \(\sim 10\) ms, instead of \(T_2\sim 2\) s in a homogeneous external magnetic field. Henceforth, the $2D$ diffusion-relaxation spectra refer to \(D-T_{2eff}\) instead of \(D-T_2\). Detailed discussion on this is given in the second section of the Supporting Information (SI). 

In the case of NMR diffusion experiments with a uniform diffusion process, the self-diffusion coefficient $D$ is obtained by appropriate fitting the $^1$H NMR spin-echo decay data \cite{Kimmich1991}. However, in non-uniform diffusion processes, diffusion is expressed with a distribution function $f(D)$, which can be obtained by implementing an appropriate inversion algorithm, as explained in the Supporting Information. The advantage of the \(D-T_{2eff}\) spectroscopy in comparison to $1D$ NMR diffusion measurements is its ability to resolve signals with different \(T_{2eff}\) and therefore acquire distinctly - otherwise overlapping - $D$ values \cite{Hassan2018}. Consequently, $D$ of the nanotube water can be resolved by analyzing the spin-echo decay signals in different $T_{2eff}$ windows.

\indent
\begin{figure*} [t]
\includegraphics[width=0.8\textwidth, keepaspectratio]{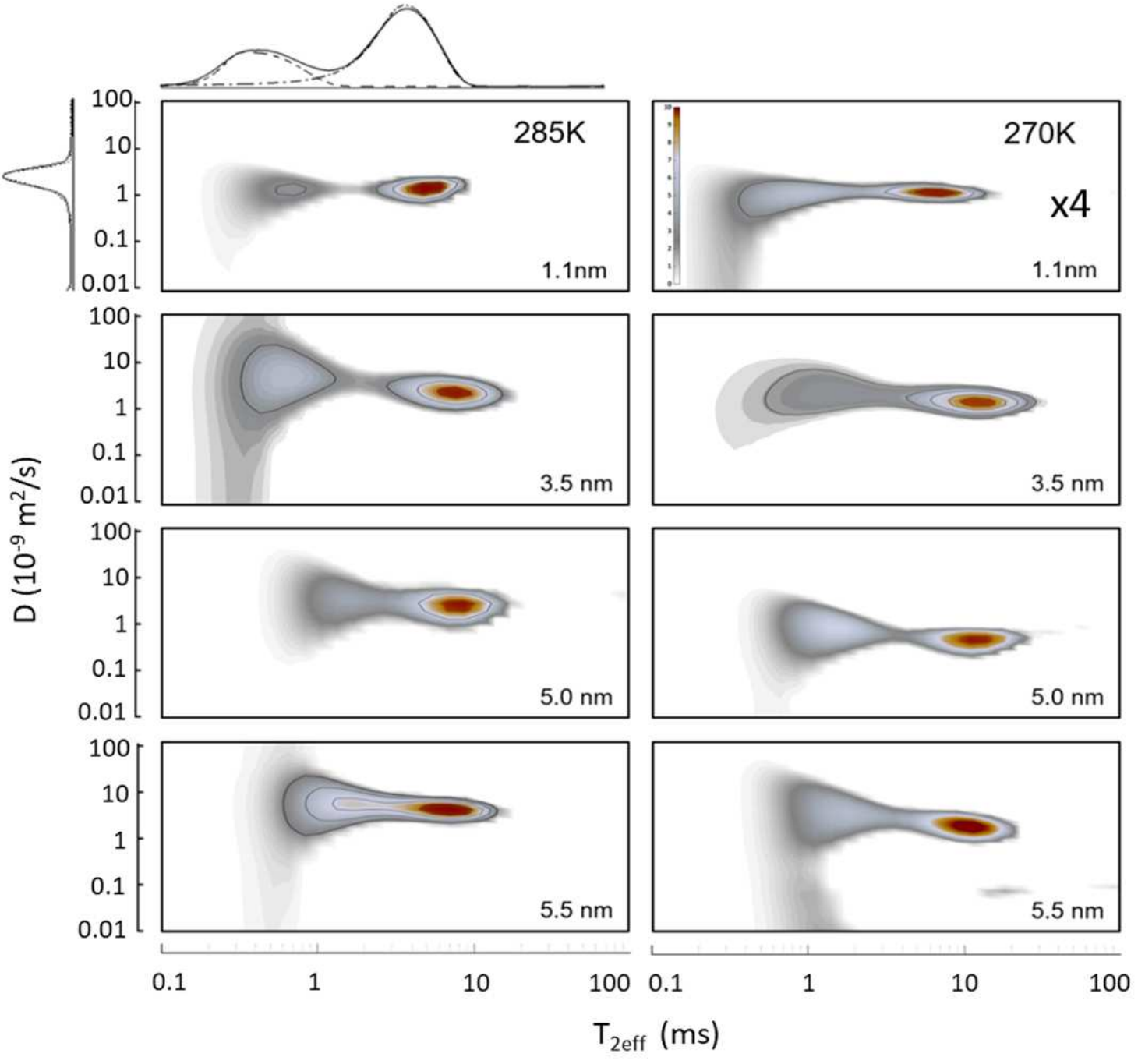}
\caption{\label{Fig2} $2D$ $^1$H NMR \(D-T_{2eff}\) contour-plots of water inside CNT of sizes $1.1$ nm, $3.5$ nm, $5.0$ nm and $5.5$ nm, at selected temperatures ($270$ K and $285$ K). Two main \(T_{2eff}\) peaks are observed, corresponding to two different water groups (interstitial and nanotube water) – as seen in the \(T_{2eff}\) projection for a $1.1$ nm sample at $285$ K. For better visualization all signal intensities at  $270$ K are multiplied by $4$.}
\end{figure*}

Figure ~\ref{Fig2} shows the $2D$ NMR \(D-T_{2eff}\) spectra of four characteristic samples measured in this study, at $270$ K and $285$ K. The intensities of the NMR contour plots are rescaled accordingly to improve visualization. Two main signals are visible acquiring different \(T_{2eff}\) values, i.e. $0.5$ ms and $10$ ms, respectively. The short \(T_{2eff})\) signal is assigned to the nanotube water \cite{Hassan2018}, while the long \(T_{2eff}\) signal is assigned to bulk and interstitial water, i.e. external water confined in the space between CNTs, which are assembled in CNT-bundles, as explained in detail in ref. \cite{Phillips2005}.   Upon lowering temperature, bulk water freezes and the intensity of the NMR signal from the external water decreases rapidly, as observed in Figure 5 of the SI. Below $273$ K bulk water becomes invisible due to the extremely low \(T_2\ \approx T_{2eff}\) of ice and only the nanotube and interstitial water are observed. 

\indent
\begin{figure}[t]
\includegraphics[height=0.8\textwidth, width=0.4\textwidth, keepaspectratio]{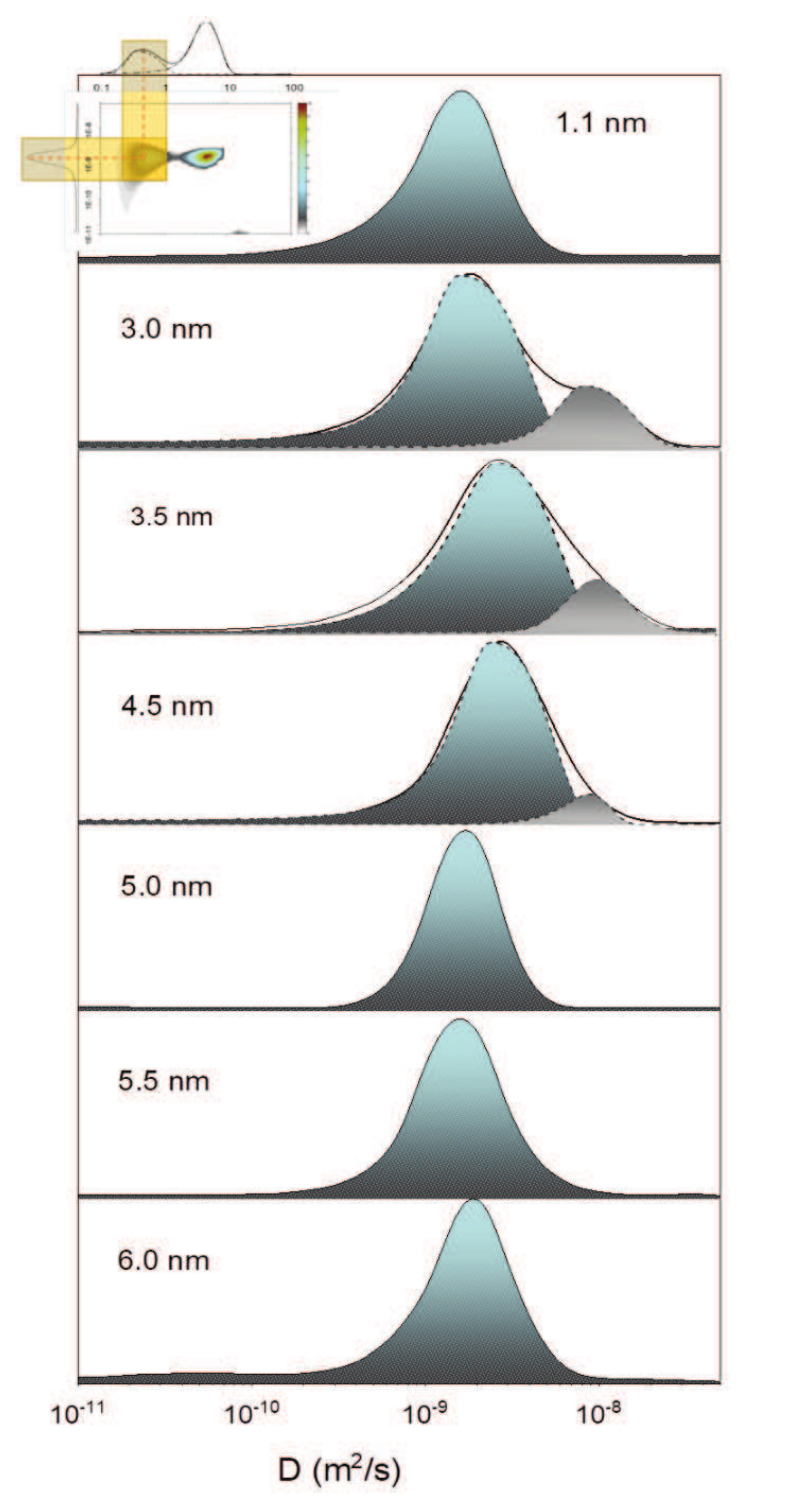}
\caption{\label{Fig3} $^1$H NMR Diffusion projections (solid lines) from the \(D-T_{2eff}\) spectra of the internal nanotube water in different CNT sizes, at $285$ K. Diffusion projections at certain CNT sizes ($3.0$ nm, $3.5$ nm and $4.5$ nm) are resolved into two components (dashed curves), represented by the main and the shoulder peaks.}
\end{figure}

\begin{figure}[t]
\includegraphics[width=0.5\textwidth, keepaspectratio]{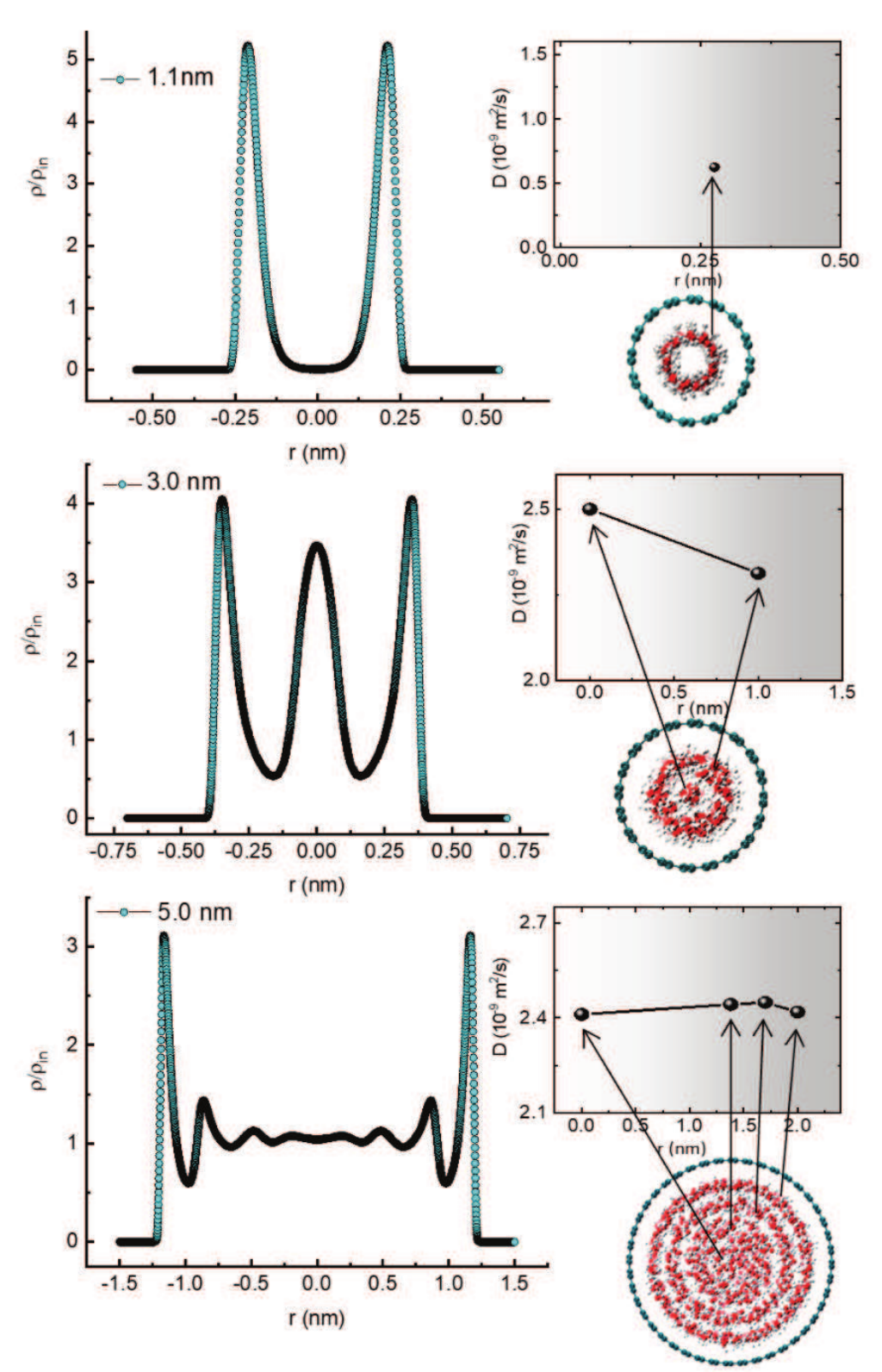}
\caption{\label{Fig4} Water local density inside different CNT sizes along with their corresponding models obtained using MD simulations, at room temperature. The x-axis is the CNT inner diameter where zero represents the center of the nanotube. Self-diffusion coefficients for the observed water layers inside CNTs were also calculated. }
\end{figure}

To uncover the nanotube water dynamics, we calculated separately the $D$ projections corresponding to the short \(T_{2eff}\) signal component, i.e. to the nanotube water. Figure ~\ref{Fig3} shows the diffusion profile of the nanotube water for all measured samples at $285$ K. In all temperatures, the diffusion curves acquire an asymmetric distribution with a long tail towards the low $D$ values, which is the fingerprint of uniform $1D$ restricted diffusion in a set of randomly oriented nanochannels \cite{Hassan2018,Callaghan1979}. Remarkably, the diffusion profiles at certain CNT sizes ($3.0$ nm, $3.5$ nm and $4.5$ nm) can be fitted with two log-norm distribution functions, with the faster one exhibiting $D$ values up to five times higher than that of bulk water. Furthermore, in the same CNT size range, the $D$ values of the main diffusion peak are sufficiently higher than those of the rest CNT sizes. For instance, it is found that \(D\sim 2.6 \times 10^{-9} m^2/s\) in the $3.5$ nm CNT, sufficiently higher than \(D\sim 1.6 \times 10^{-9} m^2/s\) in the $5.5$ nm sample. This result is in agreement with previous studies \cite{Liu2014,Ohba2013}, which show that the mean $D$ value of water in small CNT sizes is twice as large than in large CNT sizes. At larger CNT diameters $D$ acquires the value of bulk water. 
In order to understand the split of water dynamics in two components, MD simulations were conducted at room temperature in different sizes of CNTs, to reveal the local density of the water layers and the diffusion coefficients. Results are presented in Figure ~\ref{Fig4}. The stratified water arrangement is clearly seen within the CNTs. In small CNT size of $1.1$ nm, water molecules form a single tubular layer in agreement with the literature \cite{Alexiadis2008}. Due to the hydrophobic interaction between water molecules and carbon atoms of the CNT walls, water molecules of this layer are far from the CNT wall by $0.3$ nm. Furthermore, the calculated $D$ values are about \(\sim0.7 \times 10^{-9} m^2/s\) as seen in Figure ~\ref{Fig4}. 

\indent

Upon increasing the CNT diameter, additional water-layers are formed. In the $3.5$ nm size sample, MD simulations reveal two concentric WTS with different $D$ values in agreement with the NMR results. The water density profiles indicate that the outer WTS close to the CNT wall corresponds to the main diffusion peak in Figure ~\ref{Fig3}, while the central WTS corresponds to the fast diffusion component. The outer WTS shows a mean $D$ value of \(\sim 2.2 \times10^{-9} m^2/s\), in agreement with the NMR results, however, the central WTS differs significantly from the experimentally measured $D$ value at the center of the CNT channel. Finally, in the large diameter sample ($5.0$ nm), although the MD simulation has revealed multiple water layers, their calculated $D$ values are close to each other, unveiling uniform dynamics across the diameter, in agreement to the NMR results of Figure ~\ref{Fig3}. Similar analysis was performed to all samples at various temperatures. At this point, it is important to rule out the diffusion of water molecules in the radial direction of the CNT channels. This is due to the large free-energy-barrier between consecutive layers in the radial direction, which might take values as high as $1-2$ kcal/mol \cite{Alexiadis2008a}. Therefore, radial diffusion is prohibited especially at low temperatures, where molecules do not have sufficient thermal energy to overcome the free energy barrier. 

\indent
\begin{figure}[t]
\includegraphics[width=0.5\textwidth, keepaspectratio]{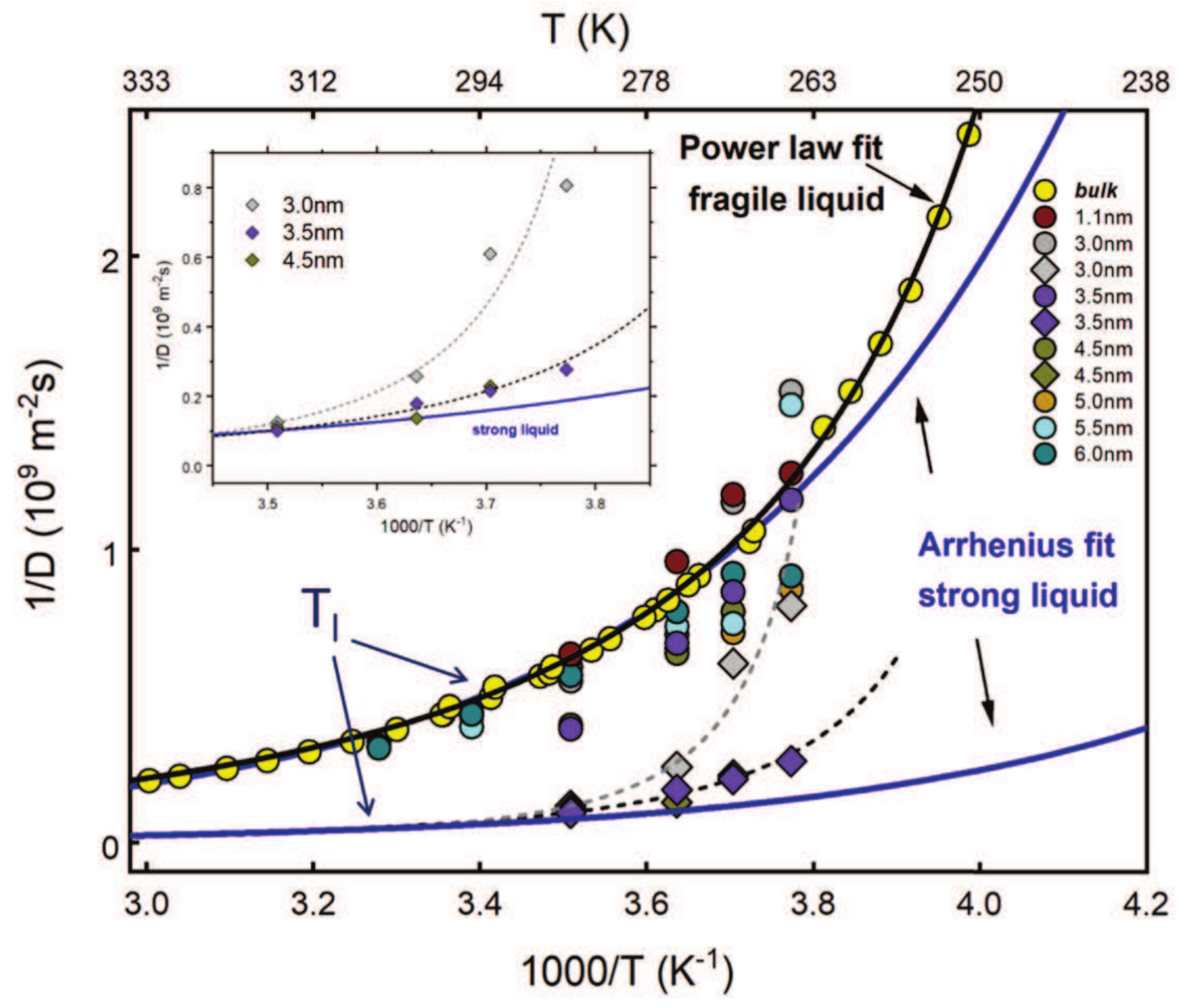}
\caption{\label{Fig5} Experimental $1/D$ vs. $1000/T$ of the nanotube water in CNTs of various sizes. The blue lines (in both the main figure and the inset) are theoretical $1/D$ vs. $1000/T$ curves of an ideal “strong” liquid obeying the Arrhenius law. The yellow circles and the black line are the experimental values of bulk water and the relevant power law fit. In CNT sizes $3.0$, $3.5$, and $4.5$ nm, two water groups are resolved with different dynamics (slow and fast). The grey and the black dashed-lines are the power line fits of the data of the fast nanotube water group. The blue arrows are the relevant liquidus temperatures \(T_l\). The inset is magnification of the $1/D$ vs. $1000/T$ curves of the fast water component for CNT sizes $3.0$, $3.5$, and $4.5$ nm.  The self-diffusion data of bulk water at temperatures below 260 K were taken from ref. 7 of the SI.}
\end{figure}

Figure ~\ref{Fig5} shows the temperature dependence of the inverse of the self-diffusion coefficient $1/D$ vs. $1000/T$, of the nanotube water in CNTs with diameter ranging from $1.1$ nm to $6.0$ nm, in the temperature range of $265$ K to $305$ K. The blue lines are the relevant curves of an ideal liquid obeying the Arrhenius law \( \frac{1}D = \frac{1} D_0 \exp\left(\frac{U}{k_BT}\right)\) for two different initial \(1/D_0\) values. Such liquids are denoted in the literature as strong liquids \cite{Shi2018}. The yellow circles are the experimental values of bulk water; at high temperatures water follows the Arrhenius law, however as shown in Figure ~\ref{Fig5}, below the liquidus temperature \(T_l \approx 273\) K, i.e. the temperature above which a material is completely liquid, strong deviation from the Arrhenius law is observed, while by approaching the glass forming transition temperature, $1/D$ obeys the Arrhenius behavior again \cite{Chu2007}. The high temperature Arrhenius and non-Arrhenius dynamic crossover at the liquidus temperature $T_l$ has been observed in many glass-forming systems \cite{Wen2017}. Liquids with this kind of behavior are denoted as “fragile” liquids. Similar to the bulk water, the temperature dependence of the diffusion coefficient of nanotube water shows strongly non-Arrhenius behavior. 
Many theoretical explanations have been proposed to explain the fragile behavior of water, such as the change in the translational and reorientation dynamics \cite{Rozmanov2017}, the coexistence of high and low-density liquid structures \cite{Xu2005,Huang2009}, the increasingly collective character of water motions at low temperatures \cite{Nicodemus2011}, the freezing of some collective motions \cite{Gallo1996,Swenson2010}, and a connection of hydrogen-bond exchange dynamics to local structural fluctuations \cite{Stirnemann2012}. For a quantitative description of our data we adopted the Speedy-Angell power-law approach, having the following form \cite{Speedy1976},
\begin{equation*}
\frac{1}{D} = \frac{1} D_0 \exp\left[\frac{T}{T_S}-1\right]^{\gamma} 
\end{equation*}
where $T_S$ is the thermodynamic limit at which transport properties become zero \cite{Ngai2011}, and the exponent $γ$ is associated with the fragility and the formation of an open hydrogen-bond network. \\
\indent
In the case of bulk water, the solid black line is the power-law fit to the experimental data (yellow circles) with \(T_S=218\) K and \(\gamma=-2\) in agreement with previously reported values \cite{Perakis2011}. Evidently, the adopted power law describes adequately the dynamical behavior of bulk water. In small and large CNT sizes ($1.1$ nm, $5.0$ nm, $5.5$ nm and $6.0$ nm), the nanotube water dynamics is similar to that of bulk water, while in the CNT size range $3.0$ nm - $4.5$ nm, two nanotube water groups are resolved with different dynamics, i.e. an outer slow component (circle data points) and a central fast one (rhombus data points). The data of the fast nanotube water group (rhombus) can be fitted to the Speedy-Angell power-law only when \(\gamma\) values in the range of $-2.0$ to $-5.0$ are considered. This indicates that the fast axial water component attains a very fragile structure \cite{Hassan2018}, resisting the formation of a hydrogen bonding network upon cooling. Besides, the liquidus temperature $T_l$ of the fast water component shifts to higher temperatures, a behavior associated with the size dependent rise of the melting temperature in very narrow single walled CNTs \cite{Maniwa2005,Pugliese2017} . 
It is furthermore observed that in the specific size range ($3.0$ nm - $4.5$ nm), the $1/D$ vs. $T$ curve of the slow water component deviates from that of bulk water; specifically, even the outer WTS close to the CNT walls acquires higher liquidus temperature and higher fragility than bulk water. This assignment differs from the picture conveyed by certain MD simulation reports \cite{Barati2011} where water close to the CNT walls is shown to diffuse faster due to pure hydrophobic interactions between the water and the CNT walls, a fact however which does not explain the dynamic spatial heterogeneity between the central and the outer WTS components revealed by our experiments.
 
\begin{figure}[t]
\includegraphics[width=0.5\textwidth, keepaspectratio]{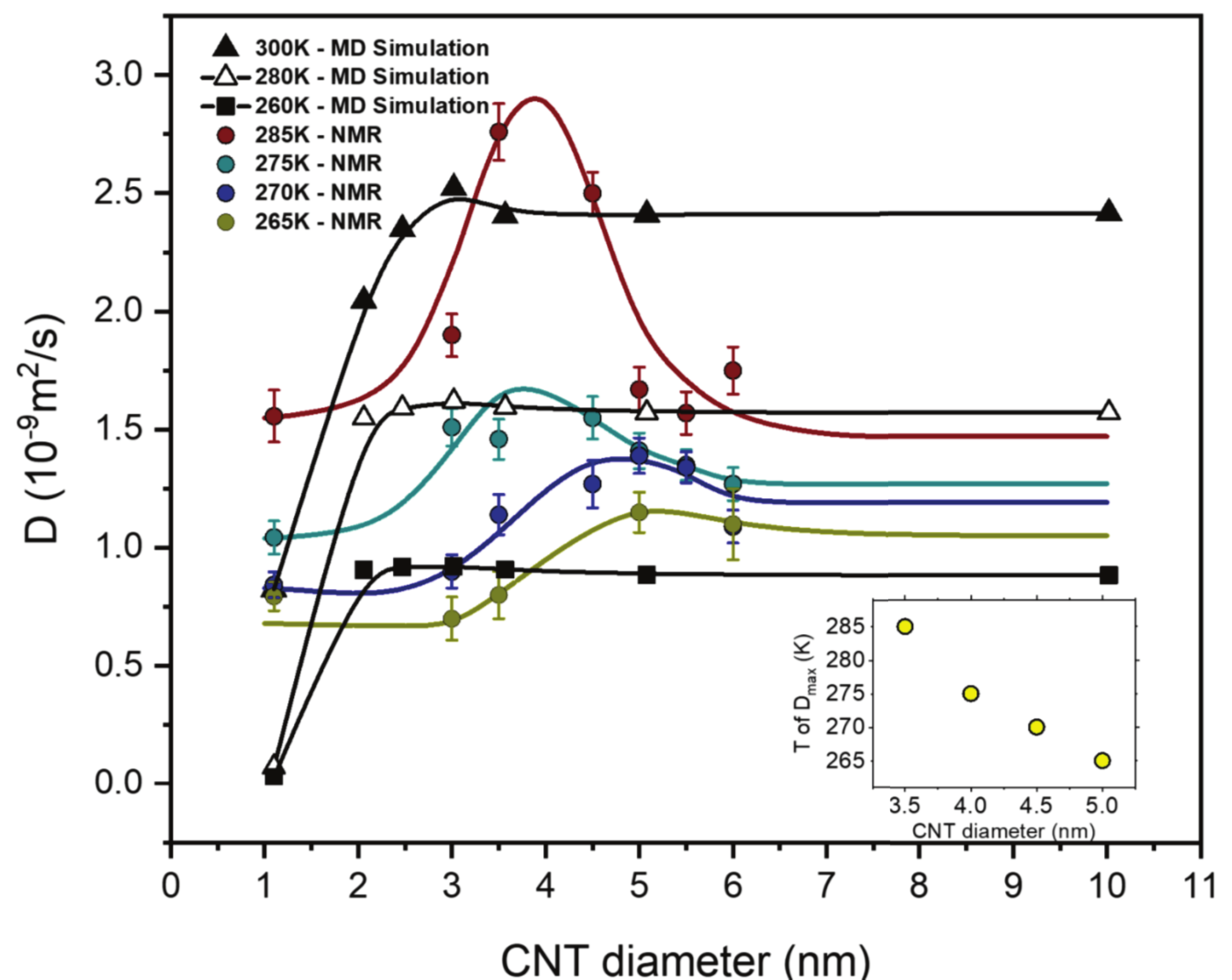}
\caption{\label{Fig6} NMR experimental self-Diffusion coefficients of the outer WTS versus CNT sizes, at different temperatures ($285$ K, $275$ K, $270$ K and $265$ K). Black colored solid triangles, empty triangles, and solid squares are MD simulation results at $300$ K, $280$ K, and $260$ K. The black solid lines are guides to the eye. The inset shows the CNT diameter of maximum diffusion (diffusion peak) at each measured temperature.  }
\end{figure}

Figure  ~\ref{Fig6} shows the CNT size dependence of the experimental $D$ values of the outer slow diffusing WTS component, together with the relevant MD simulation results.  At $285$ K, a diffusion maximum is observed in the diameter range of $3.0$ nm - $4.5$ nm, as also shown in Figure  ~\ref{Fig3}. The maximum $D$ value decreases by lowering temperature, indicating the freezing of the diffusion process. The MD simulation results (black points and black lines in Figure  ~\ref{Fig6}), are in relatively good agreement with the experimental results, however, the experimental diffusion maxima are sufficiently higher than those acquired by the MD simulations. Results in Figure  ~\ref{Fig6} confirm the prediction of diffusion maximum in ref. \cite{Barati2011}, while the smaller $D$ values  in comparison to ref. \cite{Barati2011}, acquired by our MD simulations might be due to slight differences in the boundary conditions, i.e. long open CNTs used to measure diffusion. Furthermore, the low $D$ values in the $1.0$ nm CNT is caused by the extreme confinement effect and was reported in early studies on water in small CNT sizes \cite{Mashl2003,Zuo2009}. 

\indent

Another important observation in Figure ~\ref{Fig6} is that the position of the experimental diffusion peak depends strongly on both the temperature and the CNTs size. The inset in Figure ~\ref{Fig6} shows the position of the diffusion maximum $D_{max}$, in the $T$ vs. $d$ (CNTs diameter) diagram. The highest temperature of $D_{max}$ is observed at $d =3$ nm. Evidently, this is linked with the anomalously high liquidus temperature observed in this material, which may be consider as indicator of the onset of atomic scale controlled water dynamics \cite{Pugliese2017}, and dynamically heterogeneous fragility \cite{Richert2002,DeSouza2006}. The substantial slowdown of the nanotube water dynamics by increasing the CNT size might be related to the fact that in large CNT tubes more stratified water layers are formed as also confirmed by our MD results in Figure ~\ref{Fig4}. It has been reported \cite{Alexiadis2008a} that above four or five layers the water gradually loses memory of the CNT wall and tends to acquire again the bulk water structure. Therefore, in large CNT channels, the structure of water molecules at the center of the tubes and the hydrogen bond network resemble that of the bulk water phase. 
The physical reasons behind the water diffusion enhancement in particular CNT size ($3.5$ nm) in comparison to the diffusion in larger sizes (e.g. $5.5$ nm) is still not well understood. Similar results were obtained by MD simulation \cite{Alexiadis2008,Barati2011,Holt2006} and experimental groups \cite{Liu2014,Ohba2016}. In general, two factors are expected to control the diffusion rate in various sizes of CNTs; the water structure inside as well as eventual functional groups on the CNT walls. Structural and vibrational studies have shown that water structure inside certain CNT size ($3.0$ nm) acquire  ice-like clusters \cite{Ohba2013,Ohba2014}, exhibiting cooperative motion with high diffusion. In this scheme, water clusters  can diffuse smoothly and fast into these nano-channels. The lower diffusion at smaller CNT size is due to the very small spatial restriction. On the other hand, at larger CNT size, we presume a bulk-phase water structure, due to the large available space in the inner channels of CNTs. Furthermore, previous reports \cite{Striolo2007,Majumder2011,Corry2011,Majumder2011a} suggested that the functionalization of CNTs with oxygen groups may reduce the water diffusion coefficient due to the preferential interaction between oxygenated sites and water molecules. 
To the best of our knowledge, this is the first experiment result reporting the existence of a CNT diameter range at which maximum water diffusion occurs. A similar trend is obtained by the MD simulations in Figure ~\ref{Fig6}, in agreement with previous calculations \cite{Alexiadis2008,Bordin2013,Alexiadis2008a,Barati2011,Ohba2016,Ye2011}. However, MD simulations fail to predict the experimentally detected anomalously high $D$ values of the central axial water component. It is important therefore to emphasize the difficulty to quantitatively compare between NMR and MD simulation results because: i) dynamical properties reported in MD literature are heavily depending on the water models and potential wells used \cite{Alexiadis2008a}, and ii) due to the computational power limitation, the time accessible to all MD works is in the range of ps to very few ns. On the other hand, in NMR, the accessible experimental time is typically $1-2$ orders of magnitude longer than that of MD simulations. Despite these factors, both MD simulations and NMR experiments show the same size dependence of CNTs water dynamics.
\section{Conclusions}
We have presented $2D$ NMR \(D-T_{2eff}\) results of water inside CNTs of different sizes and at various temperatures in combination with MD simulations. Our experiments show in a unique way the existence of a favorable CNT size range ($3.0$ nm – $4.5$ nm) with anomalously enhanced water diffusion. In this size range the nanotube water is further resolved into two components, with the central one exhibiting astonishing transport properties, with extraordinary high liquidus temperature $T_l$, and $D$ values ranging from two to almost four times than the $D$ values of the bulk water. Evidently, atomic scale interactions dominate water dynamics in this CNTs diameter range giving rise to the heterogeneity in the fragile behavior between the central and the outer components of the confined water. The origin of this behavior can be traced to the interrelation between the strength of the repulsive part of the interatomic potential and the liquid fragility \cite{Pueblo2017} as well as to the associated hydrogen bond lifetimes of water within the carbon nanotubes \cite{Hanasaki2006}. To the best of our knowledge, this is the first experiment result reporting on the existence of a CNT diameter range at which maximum water diffusion occurs and simultaneously exhibiting a size dependent liquid fragility. In general, the existence of new phases of water inside CNTs can add a new prospective in the field and it is an important finding on the design of nano-channels for membrane separation and drug delivery systems. 

\section{Materials and Methods}
\subsection{Materials}
Purified carbon nanotubes; Single, Double and Multiple (SWCNT, DWCNT, and MWCNT) were purchased from SES research, USA. The inner diameter of the CNTs used in this work was \(\sim 1.1\) nm for the SWCNT, \(\sim 3.5\) nm for the DWCNT and \(\sim 4.5\) nm for the MWCNT. Additionally, further MWCNT samples were purchased from Nanocs, USA with inner diameters $3.0$, $5.0$, $5.5$, and $6.0$ nm. The length of the tubes in all the CNTs used in this work was from $15$ μm to $20$ μm and all the CNTs were open ended as provided by the manufacturer. The samples were characterized using TEM-FEI Tecnai G20 with a $0.11$ nm point to point resolution and found consistent with manufacturer’s specifications. The CNT powder was used with no further treatment and doubly distilled water was used for the NMR samples preparation. Further information on the CNTs and the preparation of the NMR samples is available in the Supporting Information. 
\subsection{NMR experiments}
$2D$ $^1$H NMR diffusion-relaxation \(D-T_{2eff}\) measurements were performed in the stray field of a $4.7$ T Bruker superconductive magnet providing a $34.7$ T/m constant magnetic field gradient at $^1$H NMR frequency of $101.324$ MHz. The experiments were carried out by using a pulse sequence with more than 5000 pulses (more details can be found in the Supporting Information). The temperature was controlled by an ITC-5 temperature controller in a flow type Oxford cryostat. The accuracy of the temperature was $0.1$ K. A $30$ min time window was allowed at each temperature before collecting data. NMR data were analyzed using a $2D$ non-negative Tikhonov regularization inversion (discussed in the Supporting Information) algorithm code developed by the authors. \\
\indent
\subsection{Computational}
Molecular dynamics simulations were used to investigate the diffusion of water inside single walled carbon nanotubes. Different systems were simulated for CNT’s with different diameters. Each system consists of a nanotube of length 20nm immersed in a water bath. The nanotubes studied were armchairs \(\left(4, 4\right)\), \(\left(8, 8\right)\), \(\left(15, 15\right)\), \(\left(18, 18\right)\), \(\left(22, 22\right)\), \(\left(26, 26\right)\), \(\left(37, 37\right)\), and \(\left(73, 73\right)\) of diameters $0.55$ nm, $1.10$ nm, $2.06$ nm, $2.47$ nm, $3.02$ nm, $3.57$ nm, $5.08$ nm, and $10.02$ nm correspondingly. The molecular dynamics simulations were implemented using NAMD40. Water molecule was represented using the Simple Point Charge/Extended (SPC/E) model, which predicts accurately many of the bulk water properties \cite{Phillips2005}. The non-bonded interactions between carbon atoms were modeled using the Lennard-Jones (LJ) potential with the parameters (\(\epsilon=0.069\) kcal/mol, \(r_{min}=3.805 \AA\))  given by Werder et al \cite{Werder2003}. The positions of the carbon atoms were held fixed throughout the simulations. The systems were kept at the same temperature of $300$ K using Langevin Thermostat. In addition, the pressure was maintained at $1.0$ atm using Nosé-Hoover Langevin piston with a period of $100$ fs and a damping time scale of $50$ fs. The simulations were performed using periodic boundary conditions in which electrostatic interactions were calculated using Particle Mesh Ewald (PME). The simulation integration time step was $2.0$ fs. Bonded interactions were calculated every time step while non-bonded interaction was calculated every two steps with a cut-off of \(12 \AA\) and switching function of \(10 \AA\).
All simulated systems were minimized for 10,000 steps and then gradually heated to the target temperature of $300$ K. Each system was then equilibrated at $300$ K for $50,000$ steps ($100$ ps) before the production runs. The production simulations were run for a total time of $50$ ns. The system configuration was saved every $500$ steps ($1.0$ ps) for analysis. The water density profile was calculated for each simulated system inside the CNTs in order to elucidate the structure. 
The self-diffusion coefficient was determined using the mean squared displacement function (MSD) in the axial direction. MSD is calculated over a time interval of $1.0$ ns at a sampling rate of $1$ ps. MSD was then averaged over $50$ such time intervals. The interval length, $1.0$ ns, was chosen carefully to give water molecule enough time inside the carbon nanotube before exiting. In order to estimate the diffusion constant $D$, we fit the later part of MSD function \(\left(MSD_{t\rightarrow \infty}=2Dt\right)\) to a straight line using simple least squares regression model \(\left(\beta_0+\beta_1t\right)\). The estimated slope \(\left(\hat{\beta_1}\right)\) is equal to two times the diffusion coefficient ($2D$). In order to estimate the error in $D$ (\(\sim 10^{-12} m^2/s\)), we used the estimated standard error of the slope \(\hat{\beta_1}=\frac{\sqrt{MSE}}{S_{xx}} \). The diffusion coefficients were calculated for all of the water inside the CNT and for all the components obtained from the density profile. \\
\indent
\section{ACKNOWLEDGEMENTS}
This work was supported by Khalifa University Fund (210065). G.P. and H.J.K. would like also to express their gratitude to IRSES FP7 project Nanomag (295190). 
  \raggedright
\bibliographystyle{unsrt}
\bibliography{refcnt}

\end{document}


\preprint{SUPPORTING INFORMATION}  
\title{The Peculiar Size and Temperature Dependence of Water Diffusion in Carbon Nanotubes studied with 2D NMR Diffusion-Relaxation (\(D-T_{2eff}\)) Spectroscopy}

\author{L. Gkoura}
\affiliation{Institute of Nanoscience and Nanotechnology, NCSR Demokritos, 15310 Aghia Paraskevi, Attiki, Greece}
\author{G. Diamantopoulos}
\affiliation{Institute of Nanoscience and Nanotechnology, NCSR Demokritos, 15310 Aghia Paraskevi, Attiki, Greece}
\affiliation{Department of Physics, Khalifa University of Science and Technology, 127788, Abu Dhabi, UAE}
\author{M. Fardis}
\affiliation{Institute of Nanoscience and Nanotechnology, NCSR Demokritos, 15310 Aghia Paraskevi, Attiki, Greece}
\author{D. Homouz}
\affiliation{Department of Physics, Khalifa University of Science and Technology, 127788, Abu Dhabi, UAE}
\affiliation{Department of Physics, University of Houston, Houston TX, USA}
\affiliation{Center for Theoretical Biological Physics, Rice University, Houston TX, USA}
\author{S. Alhassan}
\affiliation{Department of Chemical Engineering, Khalifa University of Science and Technology, 127788, Abu Dhabi, UAE}
\author{M. Beazi-Katsioti}
\affiliation{School of Chemical Engineering, National Technical University of Athens, Athens, 15780 Zografou, Athens, Greece}
\author{M. Karagianni}
\affiliation{Institute of Nanoscience and Nanotechnology, NCSR Demokritos, 15310 Aghia Paraskevi, Attiki, Greece}
\author{A. Anastasiou}
\affiliation{Institute of Nanoscience and Nanotechnology, NCSR Demokritos, 15310 Aghia Paraskevi, Attiki, Greece}
\author{G. Romanos}
\affiliation{Institute of Nanoscience and Nanotechnology, NCSR Demokritos, 15310 Aghia Paraskevi, Attiki, Greece}
\author{J. Hassan}
\email[Corresponding author: ]{jamal.hassan@ku.ac.ae }
\affiliation{Department of Physics, Khalifa University of Science and Technology, 127788, Abu Dhabi, UAE}
\author{G. Papavassiliou}
\email[Corresponding author: ]{g.papavassiliou@inn.demokritos.gr}
\affiliation{Institute of Nanoscience and Nanotechnology, NCSR Demokritos, 15310 Aghia Paraskevi, Attiki, Greece}

\pacs{88.30.rh, 89.40.Cc, 66.10.cg, 68.35.Fx, 82.56.Fk, 47.56.+r}
\keywords{carbon nanotubes, water transportation, self-diffusion in liquids (mass diffusion), diffusion, multidimensional NMR, flows through porous media}

\date{\today}

\maketitle

In this document, we provide supporting information regarding the following: \\
\indent 
1. 	Materials and Methods.\\ 
\indent 
2. 	$2$D $^1$H NMR \(D-T_{2eff}\) measurements.\\
\indent 
3.	Comparison between simulated and Experimental $^1$H NMR diffusion measurements.\\
\indent 
4.	\(T_{2eff}\) projections at different temperatures.\\
\indent 
5.	Analysis of strong and fragile liquids.\\
\indent 

\section{1. Materials and Methods}
\subsection{1.1 Materials}
Purified single wall, double wall and multiwall carbon nanotubes (SWCNT, DWCNT, and MWCNT) of different sizes were purchased from SES research, USA. The inner diameter of the CNTs was \(\approx 1.1\) nm for the SWCNT, \(\approx 3.5\) nm for the DWCNT and $4.5$ nm for the MWCNT.  Further MWCNT samples were purchased from Nanocs, USA with inner diameters of $3.0$ nm, $5.0$ nm, $5.5$ nm, and $6.0$ nm. The length of the CNT channels were about $20$ $\mu m$ and all the CNTs were open ended as provided by the manufacturer. 
\subsection{1.2 Methods}
\begin{figure}[h]
          \centering
	\includegraphics[width=0.6\textwidth]{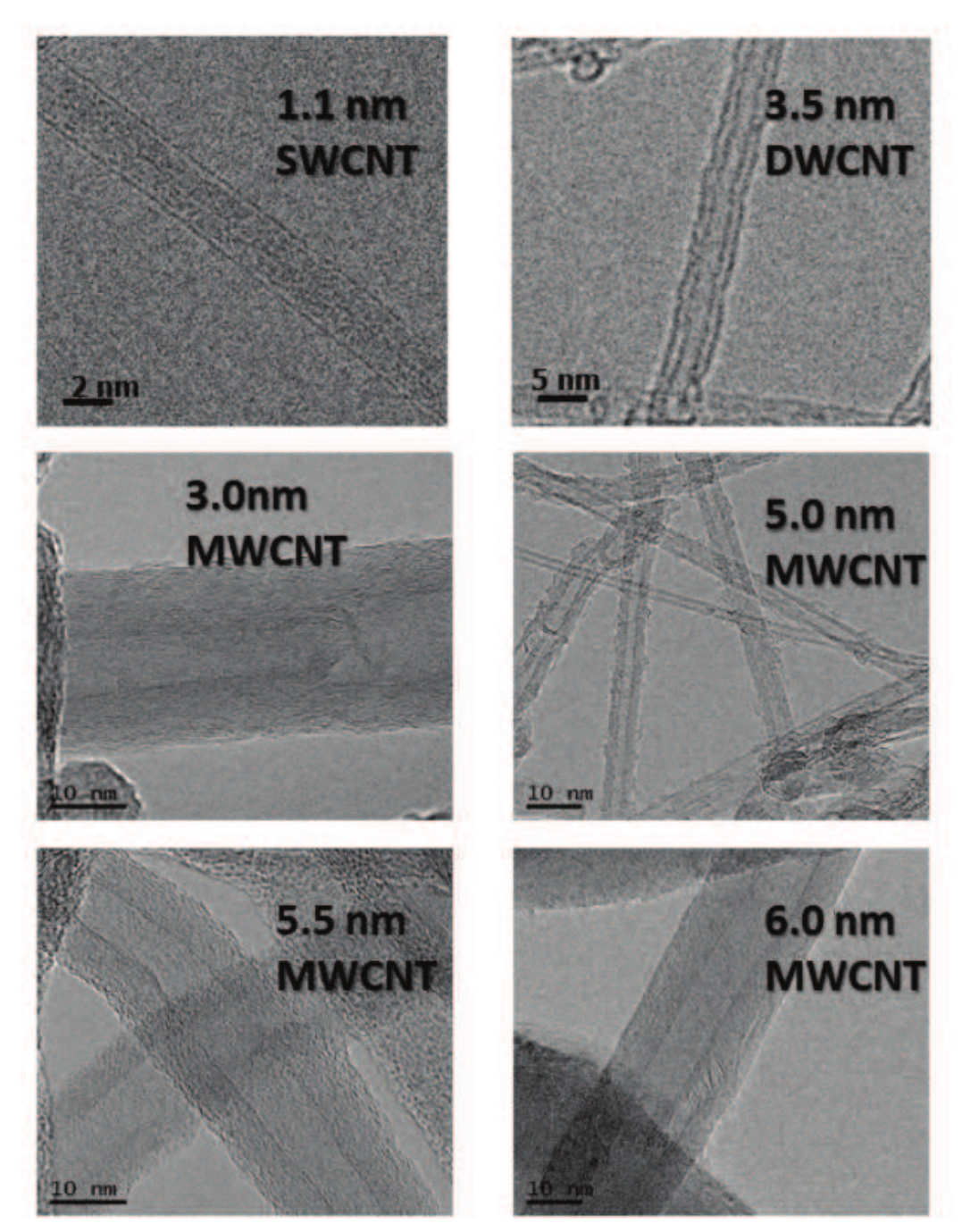}
	\caption{Representative TEM images of the CNT samples used in this study. The inner diameter of CNTs is recognizable within each image.}\label{FigS1} 
\end{figure}

\begin{figure}[t]
          \centering
	\includegraphics[width=0.8\textwidth]{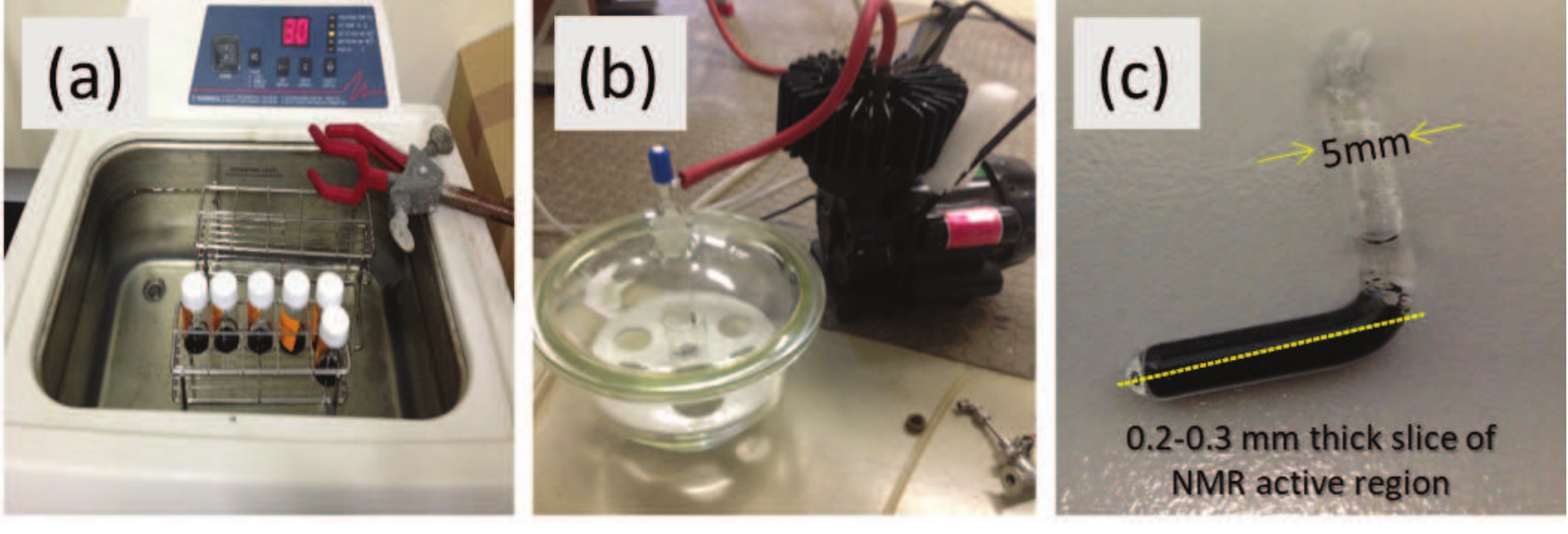}
	\caption{Preparation of CNT samples for NMR measurements. (a) Sonication of the samples: CNT powder immersed in water, (b) Vacuum system used in order to force water to replace air inside CNTs. (c) Gamma shaped NMR tube filled with water saturated CNTs. The tubes are flame sealed under vacuum on top of the vertical part, while water saturated CNTs are in the horizontal part. In the $34.7$ Tesla magnetic field gradient the NMR “active region” is a $0.2-0.3$ mm thick slice shown with yellow dashed line.}\label{FigS2} 
\end{figure}

Morphological analysis of the CNTs was conducted with Transmission Electron Microscopy (TEM) using a FEI Tecnai G20 microscope with a $0.11$ nm point to point resolution, operated at $200$ kV. For this analysis, approximately 1 mg of each sample was dispersed in $20$ ml of high purity cyclohexane (Merck, $99.9 \%$) via sonication for duration of $15$ s; a drop of each suspension was deposited on copper grids ($400$ mesh) covered with a thin amorphous carbon film (lacey carbon). To avoid contamination, CNT samples were inserted into the TEM machine immediately following preparation. Bright field images were collected at several magnifications in order to observe structure and size homogeneity. Figure S~\ref{FigS1} shows representative TEM images for the CNT samples used in this study.
To prepare fully water saturated samples for NMR measurements, CNT powder was immersed in deionized water inside gamma shaped NMR quartz tubes, with outer/inner diameters of $5/3$ mm respectively. The samples were then let to come to equilibrium at room temperature for three hours. In order to force water into the tube (expelling air out of them), the quartz tubes with the samples were connected to a vacuum system and air was carefully pumped out until small bubbles were observed leaving the immersed powder through the water, clear indication of water replacing the air inside the CNTs. The process was continued until no further air bubbles were observed. At the end of the pumping process, a water loss between $5$ $\mu l$ to $10$ $\mu l$ was observed in all water immersed samples. Each sample was then flame sealed under vacuum with a small layer of free water left in the vertical part of the tubes (Figure S~\ref{FigS2}c). 

\indent
\indent

\section{2. $2$D $^1$H NMR $D-T_{2eff}$ measurements}
\subsection{2.1 Experimental set-up}
The two-dimensional diffusion, spin-spin relaxation $2D$  \(D-T_{2eff}\) experiments, were performed in the fringe field of a $4.7$ T Bruker superconducting magnet that provides a $34.7$ T/m constant field gradient at $^1$H NMR frequency of $101.324$ MHz. For this reason the cryostat was lifted  at the position of the fringe field of the magnet, where  the NMR frequency varies linearly with the position (i.e. the magnetic field gradient in the z-direction is constant). The NMR “active” region of the water saturated CNTs in the gamma shaped tubes in the $34.7$ T/m magnetic field gradient is a horizontal slab of thickness \(\sim 0.2-0.3\) mm (Figure S~\ref{FigS2}c). The probe position was calibrated to acquire measurement at the center of the horizontal part of the tube. All measurements were performed on a broadband home-built NMR spectrometer operating in the frequency range $5$ MHz - $1$ GHz. 

\begin{figure}[t]
          \centering
	\includegraphics[width=0.8\textwidth]{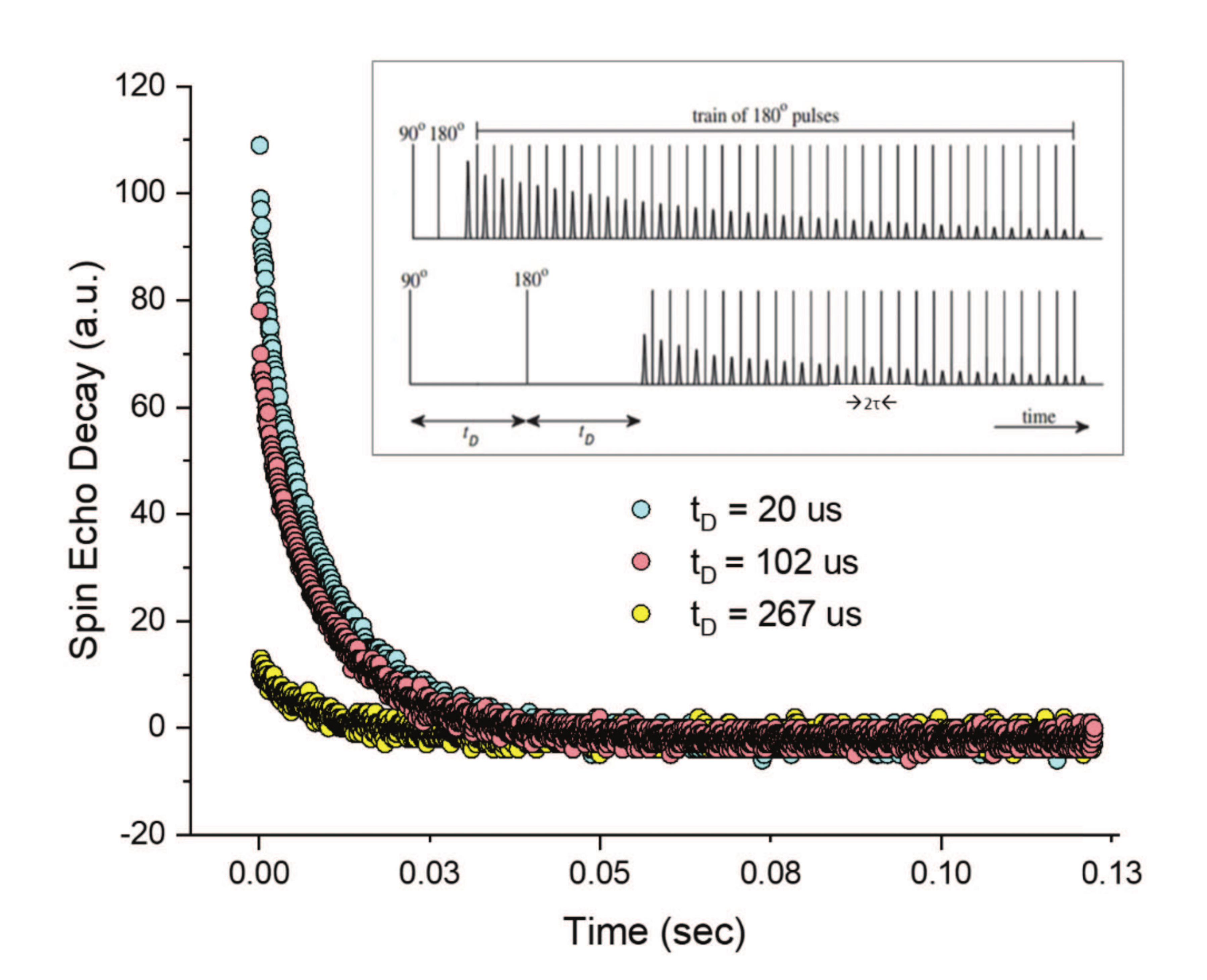}
	\caption{Upper panel shows the $2D$ \(D-T_{2eff}\) NMR pulse sequence used in this study. Lower panel shows samples of CPMG spin echo decay of water inside $3.0$ nm CNT, at $290$ K for three different \(t_D\) values. For each sample, at each temperature, around fifty values of \(t_D\) were used generating fifty similar curves as shown.}\label{FigS3}
\end{figure}
\indent
\subsection{2.2 Pulse sequence}
The pulse sequence for $2D$ \(D-T_{2eff}\) measurements is shown in the upper panel of Figure S~\ref{FigS3}. The diffusion ($D$) part of the pulse sequence is encoded in a \(\pi/2- t_D-\pi-t_D\) pulse sequence (Hahn echo) where \(t_D\) is the variable delay time between the pulses, and the  \(\pi/2\) pulse is set equal to \(4 \mu s\). The \(T_{2eff}\) part of the pulse sequence consists of a Carr-Purcell-Meiboom-Gill (CPMG) [5,6] spin-echo decay (SED) train with more than $5000$ \(\pi\) pulses, as shown in the upper panel of Figure S~\ref{FigS3}. The time interval “\(2\tau\)” between successive CPMG \(\pi\) pulses was set to \(30 \mu s\). In general, for each sample and at each temperature, around fifty \(t_D\) values were chosen ranging from \(30 \mu s\) to \(1000 \mu s\) in a logarithmic scale. In most experiments the NMR signal disappears after \(300-350 \mu s\), due to the extremely strong magnetic field gradient. \(D-T_{2eff}\) data were acquired by a series of $1D$ experiments. Data in the “direct” (\(T_{2eff}\)) dimension were collected in single shot, by recording simultaneously the intensity of all CPMG echoes. Data in the “indirect” dimension ($D$) were accumulated in successive experiments by incrementing \(t_D\) as mentioned above. 
\subsection{2.3 Signal analysis}
The recorded \(D-T_{2eff}\) signals comprise a 2D matrix \(s(t_D,t)\) decaying according to the following equation \cite{Hurlimann2002,Leu2005,Eurydice2015,Eurydice2014}:
\begin{equation} 
\frac{s(t_D,t)}{s(0,0)}=\int\int^{+\infty}_{0}\{e^{-\frac{Dg^2\gamma^2 t_D^3}{12}} e^{-\frac{t_D}{T_2}}\}e^{-\frac{t}{T_{2eff}}}f(D,T_2)\;dD\;dT_2\label{equation}
\end{equation}
where \(f(D,T_2)\) is the distribution function of $D$ and $T_2$, \(\frac{1}{T_{2eff}}=\frac{1}{T_2}+\frac{Dg^2\gamma^2\tau^2}{3}\), $g$ is the magnetic field gradient and $\gamma$ is the proton gyromagnetic ratio.\\
The first two exponentials in the brackets of equation (1) describe the attenuation of the first echo, encompassing both the diffusion and the transverse relaxation \(T_2\). The third exponential term describes the echo attenuation during the CPMG pulse sequence, due to the effective transverse relaxation time \(T_{2eff}\). In order to eliminate \(T_2\) from the time evolution of the first echo (\(t+D\)), the shear transformation \(t_t=t+t_D\) \cite{Eurydice2015,Eurydice2014} was applied, resulting the following equation:
\begin{equation} 
\frac{s(t_D,t_t)}{s(0,0)}=\int\int^{+\infty}_{0}\{e^{-\frac{Dg^2\gamma^2 t_D^3}{12}} e^{+\frac{Dg^2\gamma^2\tau^2t_D}{3}}\}e^{-\frac{t_t}{T_{2eff}}}f(D,T_2)\;dD\;dT_2\label{equation}
\end{equation}

In this way, the time evolution of the first echo becomes independent of \(T_2\). For very short CPMG interpulse time interval \(2\tau\), the exponential term proportional to \(t_D\) can be neglected in comparison to the cubic \((t_D)^3\) term, transforming equation (2) to the following equation,

\begin{equation} 
\frac{s(t_D,t)}{s(0,0)}=\int\int^{+\infty}_{0}K_0 (t_D,t_t,D,T_{2eff})f(D,T_2)\;dD\;dT_2\label{equation}
\end{equation}
where the Kernel function \(K_0\) is equal to \(K_0 (t_D,t_t,D,T_{2eff})=(e^{-\frac{Dg^2\gamma^2 t_D^3}{12}}e^{-\frac{t_t}{T_{2eff}}})\).

Equation (3) is a Fredholm integral equation of the first kind, which by inversion provides the distribution function \(f(D,T_2)\). To acquire the inversion a non-negative Tikhonov regularization method was developed. Short description of the method is given below.

\subsection{2.4 Tikhonov regularization algorithm}
In the simple case of a weak external magnetic field gradient $g$ and/or small diffusion coefficient $D$, if short interpulse time intervals \(2\tau\) is considered, then \(T_{2eff} \sim T_2\) and the Kernel \(K_0 (t_D,t_t,D,T_{2eff})\) becomes separable into a diffusion  term \(K_1 (t_D,D)\) and a \(T_2\) term \((K_2 (t_t,T_2)\). The inversion is then achieved by recasting equation (3) in matrix form as, \(\tilde{s}=\tilde{K_1}\tilde{f}\tilde{K_2}^T\), and applying a non-negative Tikhonov regularization algorithm \cite{Mitchell2012,Song2002,Day2011,Berman2013}, i.e. by minimizing the expression \( \Vert \tilde{s}-\tilde{K_1}\tilde{f}\tilde{K_2}^T\Vert^2+\lambda \Vert \tilde{f}\Vert^2\), where \( \Vert ... \Vert \) is the is the \(\hat{l}_2\) Euclidean norm of the vectors and \(\lambda\) controls the Tikhonov regularization. 
In this approach, fast minimization is achieved by decomposing \(\tilde{K_1}\) and \(\tilde{K_2}\)   separately into their respective singular values (SVD), \(\tilde{K}_{1,2}=\tilde{U}_{1,2}\tilde{\Sigma}_{1,2}\tilde{V}_{1,2}^T\), where \(\tilde{\Sigma}_{1,2}\) are diagonal matrices with their singular values in descending order, and \( \tilde{U}_{1,2}\), \( \tilde{V}_{1,2}\) are the relevant unitary matrices. 

In contrast to the above case, in the presence of a very strong magnetic field gradient $g$, as in our experiments, the kernel matrix \(\tilde {K_0}\) depends on \(T_{2eff}\) and it can not be splitted into \(\tilde{K_0}=\tilde{K_1}\otimes \tilde{K_2}\).The inversion is acquired by reformulating equation (3) in one-dimension as  \(\vec{s}=\tilde{K_0}\vec{f}\), where \(\vec{s}\) and \(\vec{f}\) are vectors obtained by lexicographically ordering the relevant matrices \cite{Mitchell2012,Song2002}. Tikhonov regularization and SVD are performed on the unsplitted total Kernel matrix \(\tilde{K_0}\). The 'best-fit" (optimal) $1D$ solution  \(\vec{f}_{opt}\) is then given by  \(\vec{f}_{opt,\lambda}=\sum_i{T_i\frac{u_i^T\vec{s}}{\sigma_i}v_i}\), where the Tikhonov filter \(T_i\) is equal to \(T_i=\frac{\sigma_i^2}{\sigma_i^2+\lambda^2}\), with \(\sigma_i\) the singular values of \(\tilde{K_0}\), and \(u_i\), \(v_i\) the left and right singular vectors forming an orthonormal basis for \(\tilde{K_0}\). The optimal \(\lambda\) value in the Tikhonov regularization was obtained automatically via error estimation with the help of a bisection method \cite{Brezinski2009}. 
Unwrapping \(\vec{f}_{opt}\) back in two dimensions with an inverse lexicographic algorithm  \cite{Mitchell2012} provides the two dimensional  \(D-T_{2eff}\) spectrum. 

\section{3. Comparison between simulated and Experimental $^1$H NMR diffusion measurements}
\begin{figure}[t]
          \centering
	\includegraphics[width=0.8\textwidth]{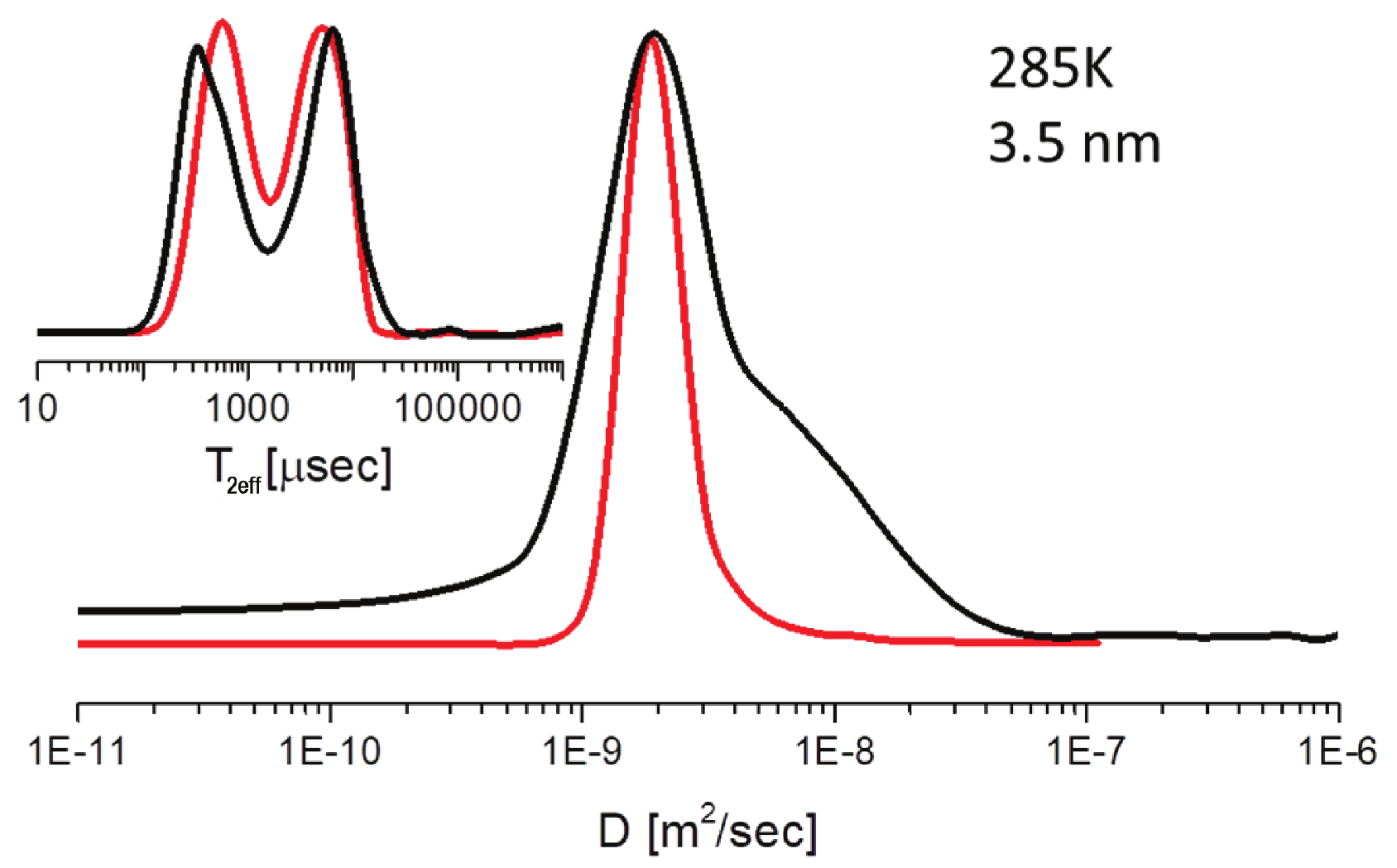}
	\caption{Comparison of the self-diffusion coefficient $D$ distribution of water in fully hydrated DWCNTs at $285$ K (black line) as obtained from \(D-T_{2eff}\) experiments with simulated data (red lines). The simulated data were acquired by setting similar \(T_{2eff}\) values with the experimental data but only one $D$-value, \(D = 1.8 \times 10^{-9}\) $m^2$/sec.}\label{FigS4}
\end{figure}

The reliability of the inversion, especially at strong magnetic field gradient and very short \(T_{2eff}\) values was checked by comparing D-profiles, acquired by experimental and simulated \(D-T_{2eff}\) data. Figure S~\ref{FigS4} presents experimental $D$ and \(T_{2eff}\)-profiles (black lines) in comparison to  simulated  ones (red lines); the latter were acquired by setting \(T_{2eff}\) values similar to the experimental ones but using a single diffusion coefficient at value \(D = 1.8 \times 10^{-9}\) $m^2$/sec. All other inversion parameters were kept exactly the same for both simulated and experimental data. 

\indent
The simulated data are observed to acquire only a slight broadening of the D-distribution (centered at \(1.8 \times 10^{-9}\) $m^2$/sec), which is an order of magnitude narrower than that of the experimental $D$-profile, showing the reliability of our algorithm. 

\section{4. \(T_{2eff}\) Projections at different Temperatures}
\begin{figure}[h]
	\includegraphics[width=0.6\textwidth]{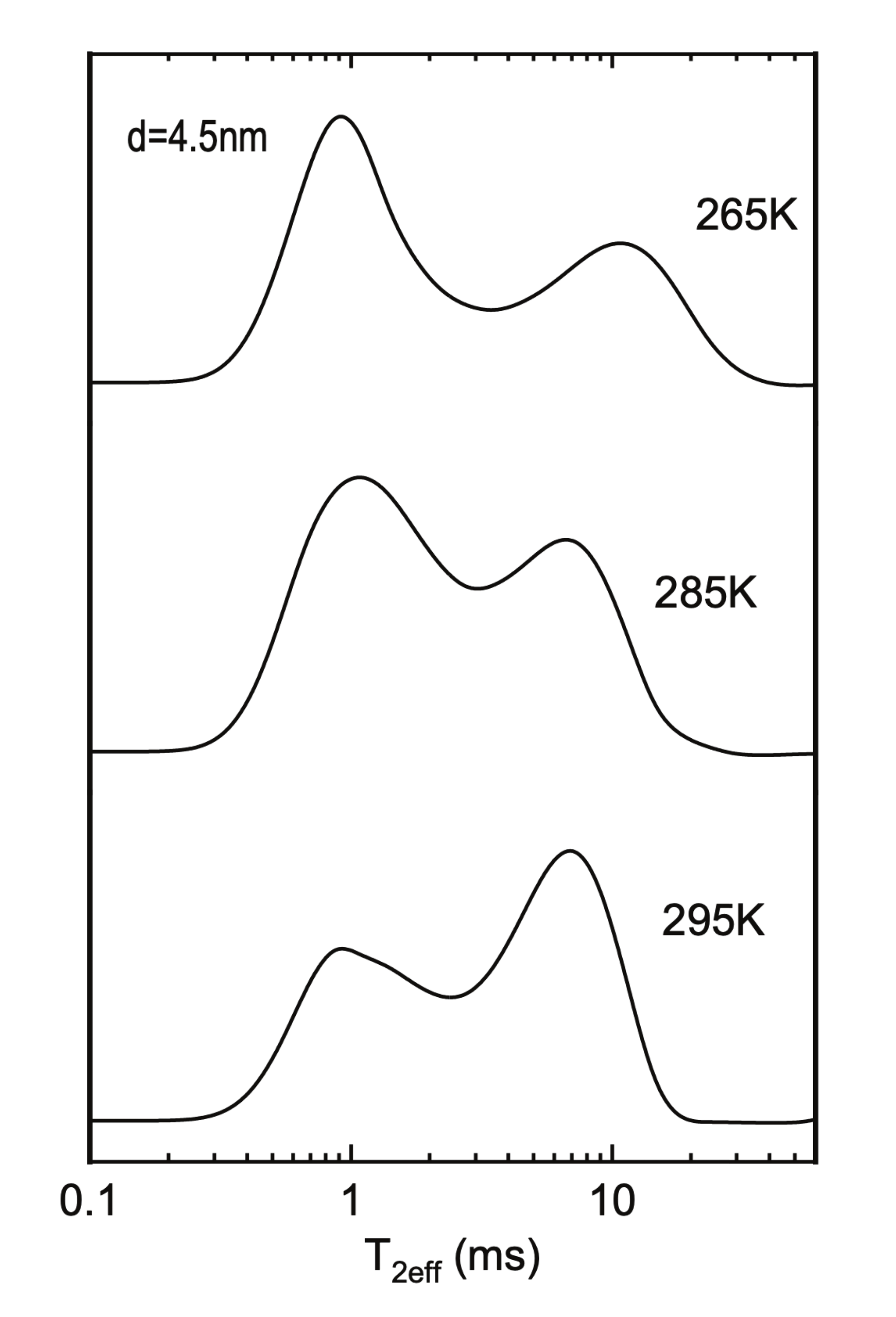}
	\caption{The $T_{2eff}$ distribution of water of the \(d=4.5\) nm CNTs, at three temperatures: $295$ K, $285$ K and $265$ K. For clarity plots have been rescaled to one. }\label{FigS5} 
\end{figure}

Figure S~\ref{FigS5} shows representative \(T_{2eff}\) projections of the $2D$ \(D-T_{2eff}\) spectra of the \(d=4.5\) nm CNTs sample at selected temperatures. Each projection exhibits two peaks, a short \(T_{2eff}\) peak corresponding to the nanotube water (water in the interior of the CNTs) and a long \(T_{2eff}\) component assigned to bulk and interstitial water among CNT bundles.  The intensity of the high-\(T_{2eff}\) signal component is shown to decrease by lowering temperature. Below the freezing temperature of bulk water ($273$ K) the long \(T_{2eff}\) component from bulk and interstitial water is assigned solely to the interstitial water, as frozen bulk water is invisible due to the very small \(T_2\) value of ice. Similar trends were observed in all measured CNTs.
\indent

\section{5. Analysis of strong and fragile liquids}
\begin{figure}[h]
          \centering
	\includegraphics[width=0.8\textwidth]{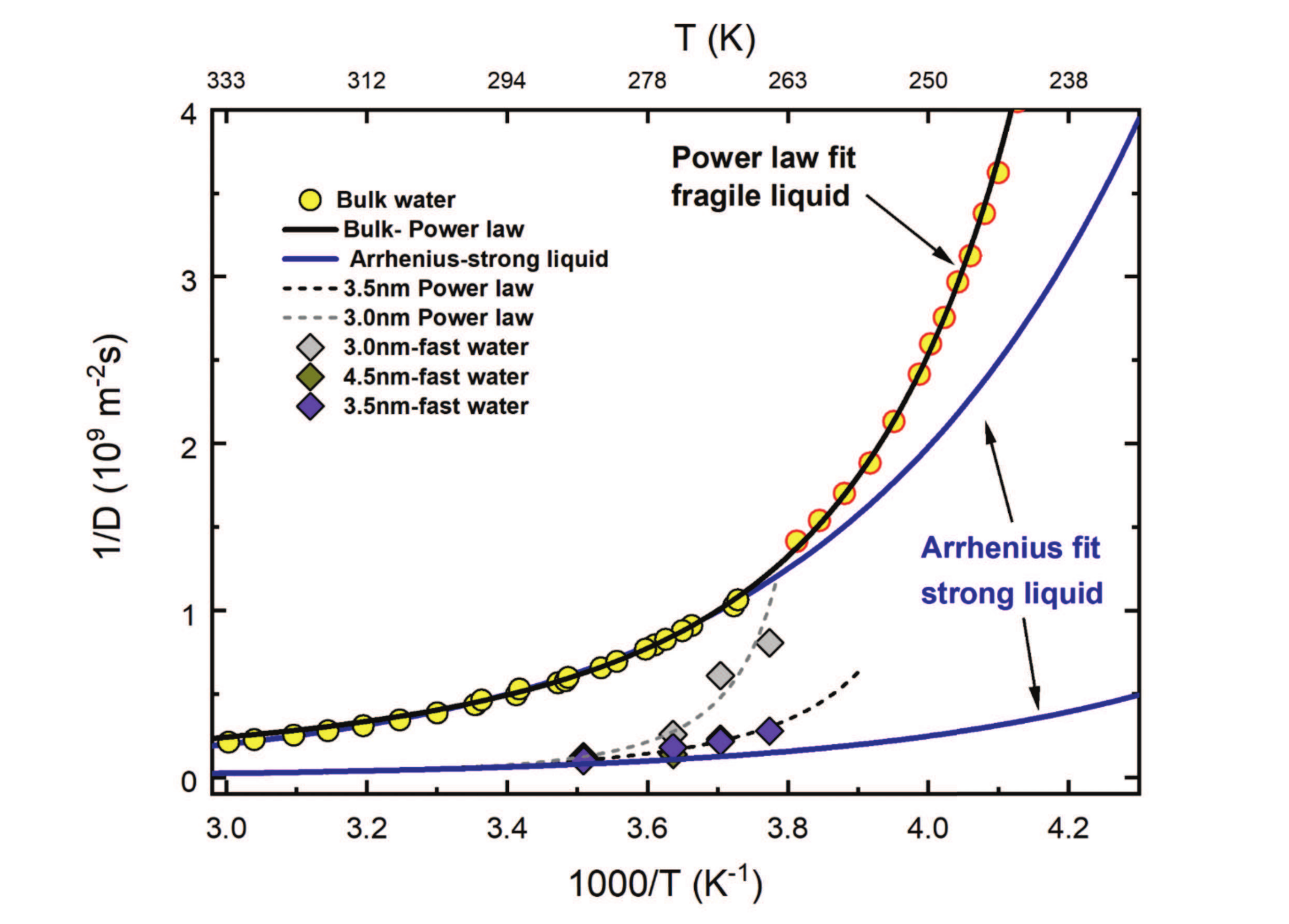}
	\caption{Temperature dependence of the inverse self-diffusion coefficient $1/D$ as a function of $1000/T$ of bulk and fast diffusive water groups inside CNTs. The black and dashed-lines are the fitting curves using the Power law. The two blue lines are the fit using Arrhenius law which do not give good fit.  }\label{FigS6} 
	\end{figure}
The temperature dependence of the inverse of the experimental self-diffusion coefficient $1/D$ as a function of $1000/T$ of bulk water and of the fast-diffusing water component inside CNTs for \(d=3\), $3.5$, and $4.5$ nm are shown in Figure S~\ref{FigS6}. The data of the bulk water shown as red-yellow circles were added from the literature (ref. \cite{Price1999,Holz2000}).The black curve was obtained by fitting the data (experimental and data from the literature) with the Speedy Angell power law \(D=D_0[(T/T_s)-1]^{\gamma}\) with \(D_0=1.635\times 10^{-9}\) $m^2$/s,  \(T_s=215.05\) K and \(\gamma=-2.063\). As seen, the experimental data deviate from the Arrhenius behavior (blue line, strong liquid) at around $270$ K, the temperature at which bulk water enters the metastable super-cooled state below the ‘’normal’’ freezing point. Data of the fast-diffusing water group inside CNTs were also successfully fitted to the Power law with fitting parameters, \(T_s=218\) K, \(\gamma=-3\) ($3.5$ nm, $4.5$ nm) and \(\gamma=-5\) ($3.0$ nm). One observation is that the cross over temperature at which data deviate from the Arrhenius behavior occurs at even higher temperature in comparison to the bulk. For the fast water group inside the $3.5$ nm this happens at around $285$ K. The degree of fragility can be estimated from the \(\gamma\) parameters; their values are higher in magnitude for the fast diffusion in comparison to that obtained for bulk water. This is clearly seen from the sudden rise of $1/D$ data points as a function of temperature in comparison with that of the bulk water. 

\raggedright	
\bibliographystyle{unsrt}
\bibliography{suppref}